\documentclass[11pt,twoside]{article}
\usepackage{asp2004}
\usepackage{psfig}
\usepackage{epsf}
\usepackage{graphics}
\usepackage{lscape}
\def\edcomment#1{\iffalse\marginpar{\raggedright\sl#1\/}\else\relax\fi}
\reversemarginpar

\begin{document}
\title{Radio behavior of four southern non-thermal O-type stars}
\author{P. Benaglia\footnote{Member of Carrera del Investigador, CONICET.} }
\affil{Instituto Argentino de Radioastronom\'\i a, C.C. 5, Villa Elisa (1894) 
Argentina and Facultad de Cs. Astron\'omicas y Geof\'{\i}sicas, UNLP,
    Paseo del Bosque S/N, 1900, La Plata, Argentina}
\author{and B. Koribalski}
\affil{Australia Telescope National Facility, CSIRO, PO Box 76, 
    Epping, NSW 1710, Australia}

\begin{abstract}
We have conducted high resolution continuum observations at 1.4 and 2.4 GHz 
with the Australia Telescope Compact Array, towards the four southern 
Of stars: CD-47\,4551, HD\,93129A, HD\,124314,
and HD\,150136. All stars have been detected at the two frequencies. 
HD\,93129A -- the only O2 I star catalogued so far, and in a double system --,
has also been observed at 17.8 and 24.5 GHz. Its radio spectrum, 
complemented with previous observations at higher frequencies, is
analyzed here.
The interpretation yields the estimate of its mass loss rate, and a non-thermal
spectral index of radiation coming from a putative colliding wind region. 
The synchrotron and corresponding inverse Compton luminosities are derived.
\end{abstract}
\thispagestyle{plain}

\section{Introduction}

Early-type stars (O to $\sim$B2) develop strong stellar
winds that strengthen while they are leaving the main sequence towards the 
Wolf-Rayet phase. The mass loss can be as high as 10$^{-5}$ 
M$_{\odot}$\,yr$^{-1}$ at velocities of thousands
of km s$^{-1}$. 
The winds are optically thick, radiate through the free-free 
mechanism, and produce an excess of flux density
from the radio to the infrared range. 
The thermal nature of the radiation can be confirmed computing a
spectral index of  about 0.6 -- 0.8 \citep{wb1975}. 
Once this radiation regime is established, 
the detection of the wind region, and consecutive 
measurement of the flux density allow a straightforward 
determination of the mass loss rate \citep{ll1993}.
However, emission characterized by a non-thermal spectral index has been
reported repeatedly from early-type stars \citep[e.g.][]{bac1989,bck2001}. 
Given the fact that the majority of these stars are not 
single, this emission is probably produced at a colliding wind region (CWR)
of a system formed by at least two early-type stars with winds.

The  particles accelerated at the CWR can be also involved in high energy
processes. Many of these stars have been detected in X-rays  
\citep[e.g.][]{po1987}, and are proposed as counterparts 
of unidentified gamma-ray sources
\citep{rbt1999}. Multifrequency studies -from radio to gamma rays-
are fundamental to completely describe the emitting regions, and
help in a comprehensive picture of the whole stellar wind phenomenon.

In the frame of a southern radio survey on early-type stars, using
the Australia Telescope Compact Array (ATCA) at maximum angular resolution, 
we have detected four non-thermal 
emitters at 3 and 6 cm \citep{bck2001,bk2004}: CD-47 4551, HD 93129A, 
HD 124314, and HD 150136. In order to  better
define the radio spectrum, we have performed ATCA observations of these
sources, at 1.4 and 2.4 GHz, and here the results are presented. 

\section{The target stars}

\noindent {\bf {HD 93129A}}. This star belongs to the Trumpler 14 cluster, 
in the Carina region. The field is very rich in early-type stars, gathering
5 of the 10 earliest catalogued stars in the Galaxy. Following 
\citet{ta2003} we adopt a distance to Tr 14 of 2.8 kpc.
HD 93129A is the unique representative of spectral type O2 If*. 
Very recently it was 
discovered to be a binary with another early-type star \citep{wa2002, ne2004},
55 mas away (or 154 AU at 2.8 kpc).
The star shows extreme values of terminal velocity ($ 3200\pm200$ km s$^{-1}$),
 effective temperature ($52,000\pm 1000$ K), and luminosity 
($\log(L/{\rm L_\odot}) = 
6.4\pm0.1$) \citep[and references therein]{ta1997}.

HD\,93129A has been detected in the radio band at 3 and 6 cm \citep{bk2004}, 
with flux densities of $S_{\rm 3cm} = 2.0 \pm 0.2$ mJy, $S_{\rm 6cm} 
= 4.1 
\pm 0.4$ mJy. The corresponding spectral index of $1.2\pm0.3$ implies 
the presence of 
non-thermal emission which, in principle, is produced at a
colliding wind region between the components of the binary system.

\smallskip

\noindent {\bf {HD 124314}}. This is an O6 V(n)((f)) field star, at a
spectro-photometric distance of 1 kpc. It has been observed 
three times at 3 and 6 cm with ATCA: on February 1998, on March 2002, 
and on April 2002, and
presented flux density variations (see \citeauthor{bck2001} \citeyear{bck2001}
for the details).
It is catalogued as a possible SB1 by \citet{gi1987}, due to 
excursions in radial velocity in excess of 35 km s$^{-1}$ that can be 
indicative
of binarity. New spectroscopic observations are needed to confirm this result.
Besides, the Washington Doble Star Catalogue lists a visual companion 
at an angular distance of 2.8 arcsec \citep{wd1997}.

\smallskip

\noindent {\bf {HD 150136}}. HD 150136 belongs to the Ara OB1 
association, at 1.4 kpc. This  
O5 IIIn(f) star has been reported as a spectroscopic binary SB2 by
\citet{ar1988}, with an O6 star as a close companion, in a 2.7 days orbit.
\citet{ma1998} listed a third component of the system, at 1.6" from the
close pair. The ATCA 3 and 6-cm former observations detected emission 
from the system at a level of $S_{\rm 3cm} = 2.61 \pm 0.03$
mJy, and $S_{\rm 6cm} = 5.57 \pm 0.03$ mJy \citep{bck2001}. 

\smallskip

\noindent {\bf {CD-47 4551}}. This field star is classified as an O4 III(f) 
by \citet{mw2003}. Its 
spectro-photometric distance was derived as 1.7 kpc (see \citeauthor{bck2001}
\citeyear{bck2001}).
There is no information of its binarity status in the literature. 
It was detected at 3 and 6 cm with ATCA, showing $S_{\rm 3cm} = 1.77 \pm 0.05$
mJy, and $S_{\rm 6cm} = 2.98 \pm 0.05$ mJy.

\section{Observations}

The observations towards the four targets were carried out with the ATCA at 
1.384, and 2.368 GHz (20 cm, and 13 cm respectively), at the array 
configuration 6A, in Dec 2003. The total
bandwidth was 128 MHz. The calibration and analysis were
performed with the \textsc{miriad} routines.
The flux density scale was calibrated using the primary calibrator
PKS B1934-638.  The resolutions achieved are about $\sim 5''$ at 13 cm, and
$\sim 7''$ at 20 cm.

Additionally, HD 93129A was observed at 17.8 and 24.5 GHz in May
2004, with the ATCA 6C array. The total bandwidth was 128 MHz. 
The flux densities were calibrated against Mars. The angular resolutions
achieved were $2.48'' \times 0.45''$ at 17.8 GHz, and $1.83'' \times 0.30''$ at 
24.5 GHz.

In both cases, ``robust'' weighted images showed the best combination of
signal-to-noise ratio and sidelobe supression. The shortest baselines were 
excluded to get rid of the diffuse emission from nearby extended sources.

\section{Results}

We detected the four targets at 1.4 and 2.4 GHz, and HD 93129A also at 17.8 and
24.5 GHz. Gaussian 
fittings show all of them as point sources. The observed flux densities 
at 1.4 ad 2.4 GHz are listed in Table 1, together with the spectral indices. 
The flux densities detected towards HD 93129A, using the 3-mm ATCA 
receiver are: 
$S_{\rm 17.8GHz} = 1.8\pm0.2$ mJy, and $S_{\rm 24.5GHz}= 1.5\pm0.2$ mJy.
The spectra are displayed in Figures 1 and 2, combined with
the results of previous observations at 4.8 and 8.64 GHz.
For all detections, the polarization factors remain below 2\%.

\begin{table}[!ht]
\caption{ATCA flux densities at 1.4 and 2.4 GHz}
\smallskip
\begin{center}
{\small
\begin{tabular}{l l r r r }
\tableline
\noalign{\smallskip}
Star & Sp. class.$\dag$& $S_{\rm 13cm}$ & $S_{\rm 20cm}$ & $\alpha_{13-20}$ \\
     &           & [mJy]         & [mJy]         & \\
\noalign{\smallskip}
\tableline
\noalign{\smallskip}
CD-47 4551 & O4 III(f)   & $3.82\pm 0.20$ & $3.80\pm0.25$ & $+0.01\pm 0.2$\\
HD 93129A  & O2 If*      & $7.58\pm 0.50$ & $9.38\pm0.50$ & $-0.40\pm 0.4$\\
HD 124314  & O6 V(n)((f))& $3.70\pm 0.20$ & $2.83\pm0.25$ & $+0.50\pm 0.2$\\
HD 150136  & O5 IIIn(f)  & $3.05\pm 0.20$ & $2.28\pm0.40$ & $+0.54\pm 0.2$\\
\noalign{\smallskip}
\tableline
\end{tabular}
}
\end{center}
$\dag$: Ma\'{\i}z-Apell\'aniz et al. 2003
\end{table}

\begin{figure}[!ht]
\vspace{1cm}
\plottwo{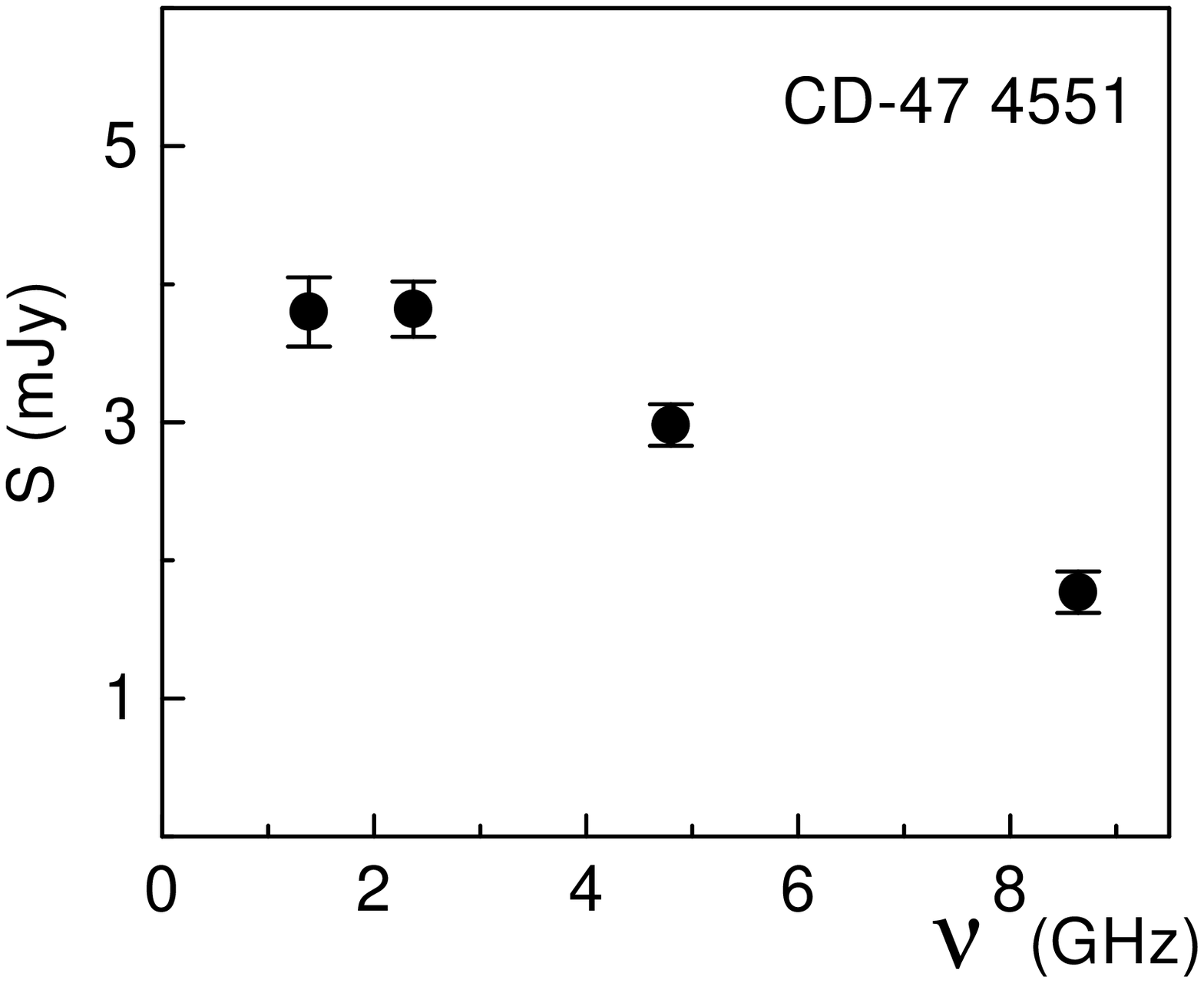}{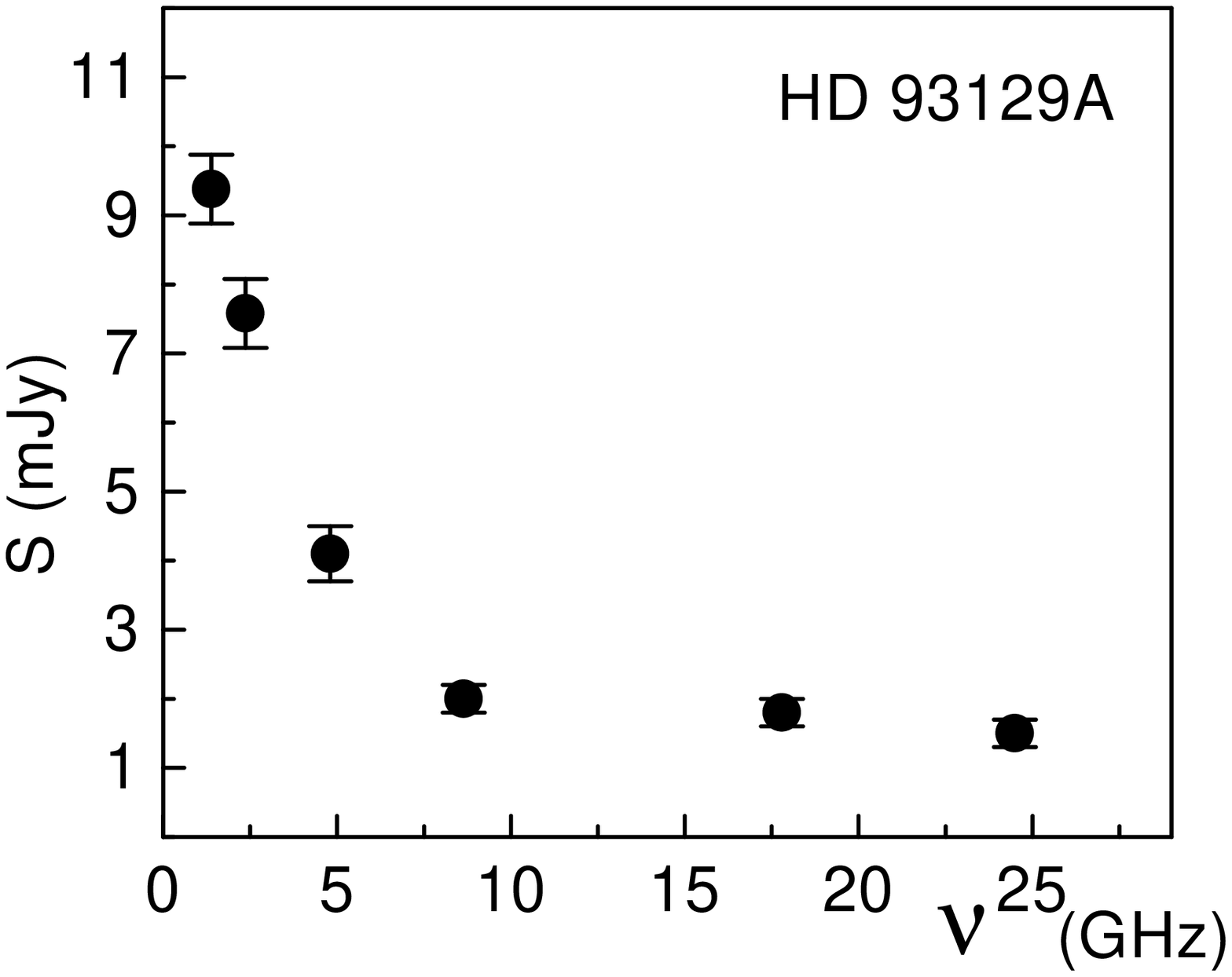}
\caption{Radio spectra of the stars CD-47 4551 and HD 93129A}
\end{figure}

\begin{figure}[!ht]
\vspace{1cm}
\plottwo{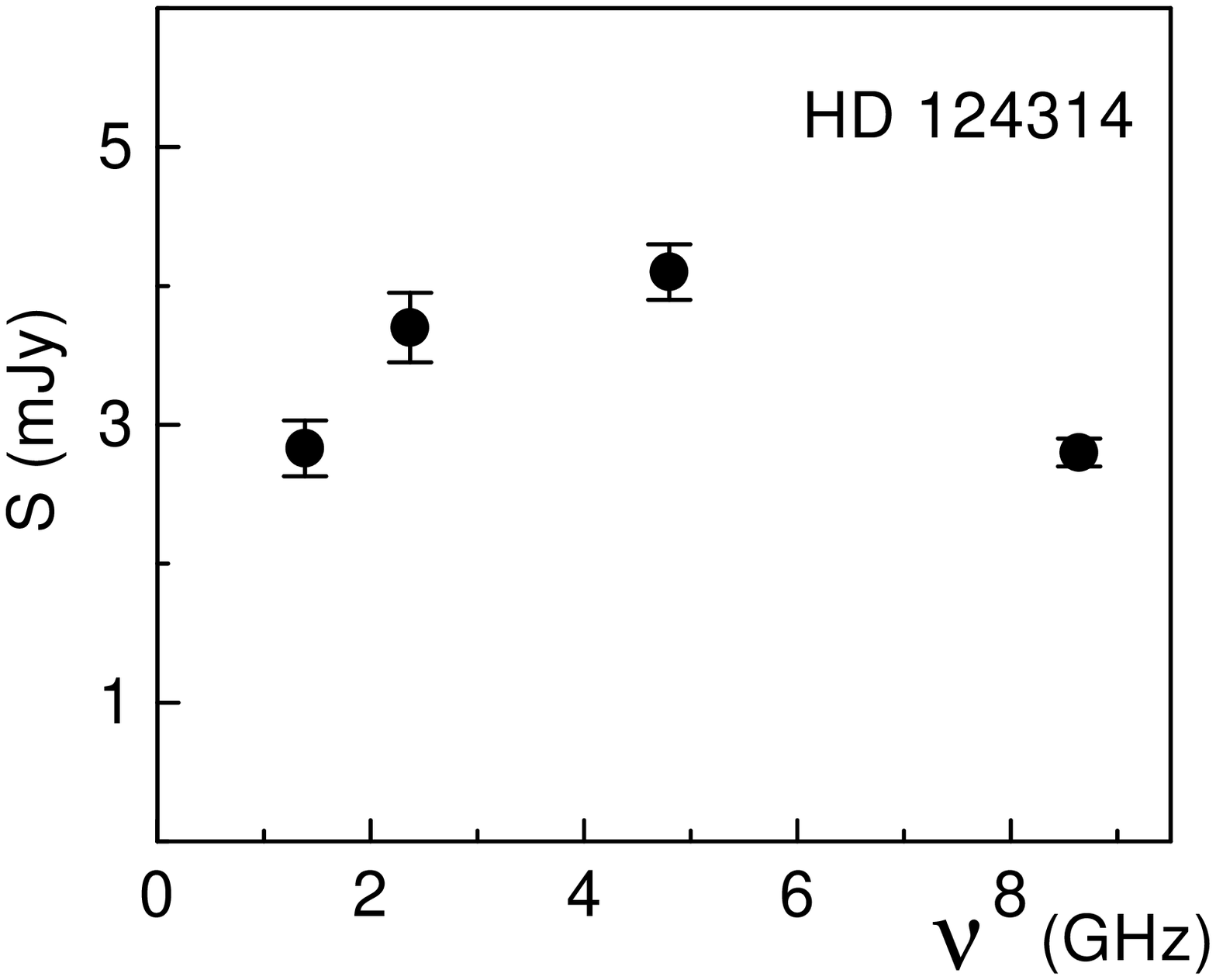}{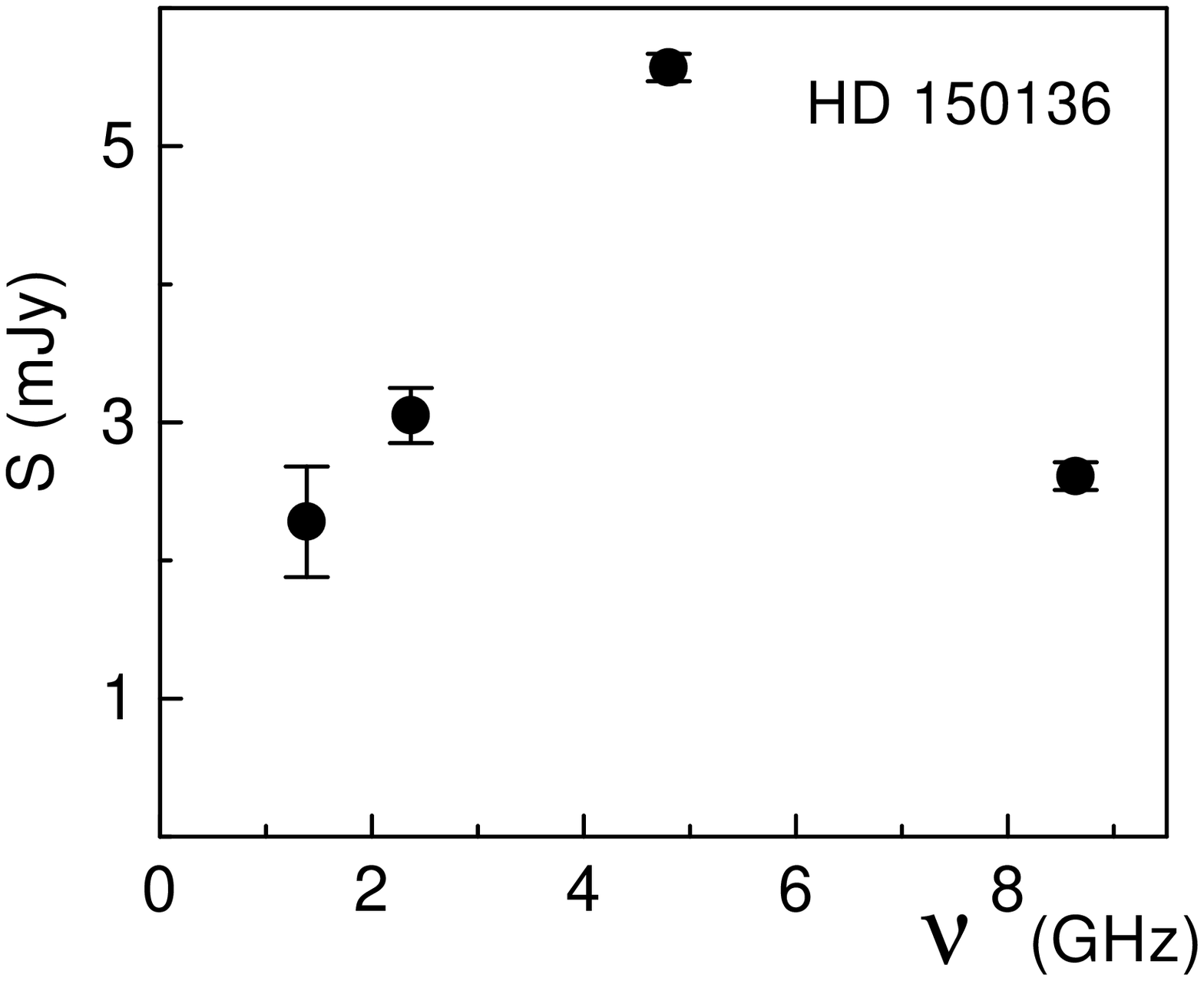}
\caption{Radio spectra of the stars HD 124314 and HD 150136}
\end{figure}

\section{The HD 93129A radio spectrum}

In close stellar systems, flux variability needs to be 
taken into account. An interpretation of the present
multifrequency radio data from CD-47 4551, HD 124314 and HD 150136 can
be envisaged only once the structure of the system is revealed. Optical
observations are currently under way to investigate this issue
(Niemela et al., in preparation).

We note that the component separation in the HD93129A system
suggests a period larger than 50 yr. Here we report radio
observations between 1.4 and 25 GHz. No flux density could be
measured in the MOST 843 MHz maps due to source contamination.

The relatively low angular resolution of the radio data does
not allow us to resolve the binary system. We detect the sum
of the thermal emission from the winds of both stars as well
as non-thermal emission from the colliding wind region (CWR).

\subsection{Spectral contributions}

The thermal emission from the winds of both components 
can be approximated with a 
radio flux density of $S_{\nu} \propto \nu^{0.6}$.
At the CWR, strong shocks are capable of accelerating wind particles 
to relativistic
energies giving rise to synchrotron emission in the presence of magnetic
fields \citep{gs1964}. However, part of the synchrotron photons 
can be absorbed by the wind thermal ions.
The contribution of synchrotron emission to the flux density, modified by
thermal absorption, can be expressed as $S_{\nu} \propto \nu^{\alpha_{\rm NT}} 
\, e^{\tau_0 \nu^{-2.1}}$.
The influence of synchrotron self-absorption (SSA) would cause the spectrum at short 
frequencies to
be represented by $S_{\nu} \propto \nu^{-2.5}$.
We will disregard the consequences of the Razin-Tsytovitch effect, because
according to the value of the magnetic fields involved, it would affect the
emission up to some MHz.

\subsection{Fitting results}

We fitted the above contributions to the spectra of Fig. 1:right, and 
found the best fit was given by 
$ S_\nu = {\rm A} \nu^{0.6} + {\rm B} \nu^{\alpha_{\rm NT}} 
e^{-\tau_0 \nu^{-2.1}}$ mJy, if $A = 0.17\pm0.05$, 
$B = 28.6\pm6.6$, $\alpha = -1.31\pm0.18$,
and $\tau_0 = 1.41\pm0.37$ (see Figure 3).
The lack of data below 1.4 GHz precluded the search for a SSA signature.

The above expression helps us to separate thermal from non-thermal emission, 
and to compute a mass loss rate from the thermal radio flux. 
At 3 cm the total flux measured is $2.0\pm0.2$ mJy \citep{bk2004}. 
Consequently, $S_{\rm T} = 0.6$ mJy, and $S_{\rm NT} = 1.4$ mJy, and
the mass loss rate for HD\,93129A results in $\dot{M} = 3.6 \times 
10^{-5}$ M$_\odot$\,yr$^{-1}$.

If the secondary star is an 03.5 V, like the nearby stars HD 93128 and 
HD 93129B, the colliding wind region would be $\sim$ 120 AU from the primary,
and $\sim$ 34 AU from its companion, in an \citet{eu1993} colliding wind
scenario.
A synchrotron luminosity of $ 6 \times 10^{33}$ erg s$^{-1}$ is obtained
up to 24.5 GHz.

\begin{figure}[!ht]
\vspace{2.5cm}
%\plotone{hd93fit.eps}
\plotfiddle{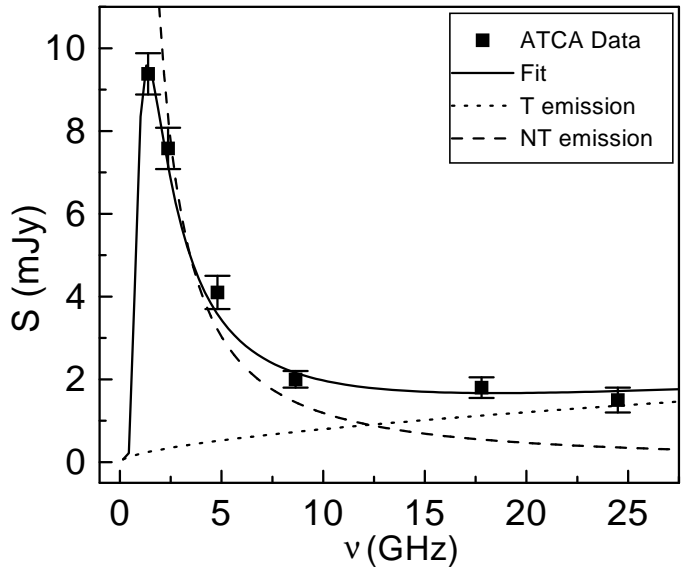}{6.5cm}{0}{55}{55}{-170}{-186}
\caption{Spectral fitting for HD 93129A}
\end{figure}

\section{Discussion}

The system HD 93129A has a separation similar to that of WR 146 (210 AU, 
\citeauthor{sg2000} \citeyear{sg2000}). 
The latter is a WR+O system that has been  resolved in the radio continuum with MERLIN, 
and the emission probably coming from
a CWR could be mapped \citep{wd2000}, as a 38 mas source (or $\sim$ 50 AU
at a distance of 1.25 kpc).
It seems reasonable to assume that the size of a colliding wind
region of HD 93129A should be alike, and thus we adopt for it a value
of 40 AU.
The equipartition magnetic field at the
CWR will result in $\sim$ 20 mGauss, \citep{mi1980}, and the stellar magnetic field, $\sim$ 500 Gauss.

The same electrons involved in synchrotron processes can be scattered
by the inverse Compton mechanism, producing gamma-ray continuum emission.
The maximum Lorentz factor attained by the electrons at the CWR, computed 
as in \citet{br2003}, is $\gamma_{\rm max} = 1.8 \times 10^{6}$. This
corresponds to 
a maximum synchrotron frequency of $\nu = 2 \times 10^{15}$ Hz, and
a cutoff energy for the IC photons of $E = 500$ GeV.
The energy at which the electron distribution changes, due to synchrotron
and inverse Compton losses, is defined by $\gamma_{\rm b} \sim$ 20,000.
Finally, the inverse Compton luminosity results in $L_{\rm IC} = 1 \times
10^{33}$ erg s$^{-1}$, well below the EGRET threshold at the location of
this particular star.

Optical spectroscopy to define the spectral type of the HD 93129A companion, 
and very high resolution radio observations to resolve the system, are 
crucial to refine the figures given here.
The gamma-ray predictions can be confronted against GLAST observations in the
near future.

In the case of CD-47 4551, HD\,124314, and HD\,150136, the 
knowledge of the system structure is fundamental to allow the 
study of the radio data.

\section{Conclusions}

Non-thermal contribution to the radio emission has been detected for the
stars CD-47\,4551, HD\,93129, HD\,124314, and HD\,150136. The spectral fitting 
for the radio emission from the O2 If* HD\,93129A enabled to extract
the thermal flux, and determine a value for the mass loss rate
of this peculiar system, of $3.6\,\times\,10^{-5}$ M$_\odot$ yr$^{-1}$.
The non-thermal radiation, coming probably from a colliding wind region
between the stellar components, is characterized by a flux density spectral
index of $-1.3\pm0.2$, and a magnetic field of $\sim$ 20 mGauss.

\acknowledgments{P.B. is grateful to G.E. Romero for useful discussions,
and wishes to thank Fundaci\'on Antorchas, FCAGLP (UNLP), and 
the University of  Montreal for financial support in attending the meeting.}

\end{document}